\documentclass[twocolumn]{revtex4}
\usepackage{amssymb}
\usepackage{latexsym}
\usepackage{epsfig}

\usepackage{amsmath}
\begin{document}
\title{ Exploring the Sharma-Mittal HDE models with different diagnostic tools }
\author{ Umesh Kumar Sharma$^{1}$\footnote{ sharma.umesh@gla.ac.in},  Vipin Chandra Dubey$^{2}$\footnote{ vipin.dubey@gla.ac.in}}
\address{$^{1,2}$ Department of Mathematics, Institute of Applied Sciences and Humanities, GLA University\\
	Mathura-281406, Uttar Pradesh, India\\
}

\begin{abstract}
	In this paper, we have examined the Sharma-Mittal holographic dark energy model (SMHDE) in the framework of an isotropic and spatially homogeneous flat Friedmann-Robertson-Walker(FRW) Universe by considering different values of parameter $\delta$ and $R$, where the infrared cut-off is taken care by the Hubble horizon. We examined the SMHDE model through the analysis of Statefinder hierarchy and the growth rate of perturbation. The evolutionary trajectories of the statefinder hierarchy $S_3^1$, $S_3^2$ $S_4^1$, $S_4^2$ versus redshift z, show satisfactory behavior throughout the Universe evaluation. One promising tool for investigating the dark energy models is the composite null diagnostic(CND) $\{ S_3^1 - \epsilon\}$, where the evolutionary trajectories of the  $S_3^1 - \epsilon$ pair present different property and the departure from $\Lambda$CDM could be well evaluated. Additionally, we investigated the dynamical analysis of the model by $\omega_{D}-\omega^{'}_{D}$ pair analysis.
\end{abstract}
\maketitle

\section{Introduction}
Various cosmological observations show that our Universe is going through accelerated expansion phase at present \cite{ref1,ref2,ref3,ref5,ref6,ref11}. The concept of dark energy (DE) was used to explain this accelerated expansion of the cosmos, where DE has negative pressure \cite{ref12,ref13,ref13a,ref13b}. There are basically two methods to explain the late-time acceleration of the Universe. Firstly, the dynamical dark energy model in which the matter part of the Einstein field equation can be changed. In all the theories and models, the cosmological constant model is the simplest one and was elucidated by Einstein \cite{ref14a,ref14b,ref14c,ref14e}, which gives resultant in the face of the equation of state parameter (EoS) $\omega = - 1$, the most basic applicant for dark energy is the cosmological constant and it is consistent from the prospective of observations, except the coincidence and fine-tuning problem  \cite{ref14b,ref14g}. As an answer to the problem various dynamical dark energy models are given as an alternative like quintessence \cite{ref14h,ref14i},  phantom \cite{ref14j},  $k$-essence \cite{ref14k,ref14k1,ref14k2}, tachyon \cite{ref14l} and Chaplygin gas \cite{ref14m}. Secondly, by modified gravity theories, which are achieved by modifying the  geometric part of Einstein field equation \cite{ref15,ref15a,ref15b,ref15c,ref15d,ref16}.\\

Many dark energy models have been proposed so far to explain the accelerated expansion phase of the Universe inspired by the holographic principle,  which in last propagates the theory that the degree of freedom is dependent on the bounding area and not on volume \cite{ref18,ref19,ref20,ref21}. M.Li, in 2004, proposed Holographic dark energy (HDE) taking future event horizon as IR cutoff to describe the accelerated expansion scenario of the Universe \cite{ref24}. The holographic dark energy model has been taken into account broadly and studied in the literature  \cite{ref23,ref17,ref25,ref26,ref27}, as $\rho _D \propto \Lambda ^4$, and the relation between the UV cutoff $\Lambda$,  entropy $S$ and IR cutoff $L$ is $\Lambda{^3} L{^3} \leq (S)^{\frac{3}{4}}$, that shows that the combination of the entropy with the IR cut-offs gives energy density of HDE model. The declaration of  $\rho_{D}$ is the focal point and is achieved by consideration of the dimensional analysis \& the holographic principle instead of the inclusion of the expression of the dark energy into the Lagrangian. This is the basis for the importance of the HDE and the original Holographic DE model is dependent  on Bekenstein-Hawking entropy $S = \frac{A}{4G}$, where $A = 4 \pi L^{2}$, so the density is $\rho_{D}= \frac{3c^{2}}{8 \pi G}L^{-2}$,  here $c$ is numerical constant. Three years after the HDE i. e. in 2007, Cai proposed the Agegrapic dark energy (ADE) model taking length measure as the age of the Universe \cite{ref28}. Due to some confusion in the original ADE model proposed by Cai, Wei and Cai in 2008, proposed the New agegraphic dark energy (NADE) model considering conformal time as time scale \cite{ref29}. The Ricci dark energy was proposed by Gao et al. \cite{ref30} replacing future event horizon with Ricci scalar curvature inspired by the holographic principle.\\

Recently, Different entropies \cite{ref30a,ref30b,ref30c,ref30d} have also been used  to propose some new forms of dark energy model in the investigation of gravitational  and the cosmological incidences. Inspired by hologrphic principle and using various system entropies, some new form of dark energy modolels were proposed, for example, the R$\acute{e}$nyi holograpic dark energy (RHDE) model \cite{ref35}, Tsallis holographic dark energy (THDE) model  \cite{ref36}, Tsallis agegraphic dark energy (TADE) model  \cite{ref37} and Sharma-Mittal holograpic dark energy (SMHDE) model  \cite{ref38}. These newly proposed dark energy models were investigated by various researchers in different scenario \cite{ref38a,ref38b,ref38c,ref38d,ref38e,ref38f,ref38g,ref38h,ref38i,ref38j,ref38k,ref38l,ref38m}.\\

As the number of dark energy models is increasing day by day, the diagnostic tools which can discriminate them are required. The statefinder hierarchy and the growth rate of linear perturbations, as null diagnostics for the $\Lambda$CDM model, was introduced by Arabsalmani and Sahni \cite{ref87} to discriminate the different dark energy models from the $\Lambda$CDM model. The statefinder hierarchy contains high derivatives of scale factor $a(t)$, model-independent and is 
a geometrical diagnostic \cite{ref88}. Previously to check scale-independent consistency between the structure growth and the expansion history, the growth rate of the structure was used in \cite{ref89,ref90,ref91}. It can be combined with the statefinder hierarchy or act as a cosmic growth history diagnostic to serve on a composite diagnostic. Four holographic DE models were discriminated against by these two diagnostics in \cite{ref88}. In \cite{ref91,ref92,ref93,ref93a,ref93b,ref93c,ref93d,ref93e,ref93f} these diagnostics were considered. Recently,  the discrimination between  THDE models  $\Lambda$CDM model investigated by one of the authors through statefinder hierarchy in the nonflat Universe considering apparent horizon as IR cutoff \cite{ref93g}. The $\omega_{D}-\omega_{D}^{'}$ analysis \cite{ref57} can also be utilized to recognize the difference in dark energy models, which is based on the behavior of EoS for the dark energy models.\\

In this work, we have explored the newly proposed  Sharma-Mittal Holographic Dark Energy (SMHDE) model through the diagnostic tools described above in the flat FRW Universe by taking the Hubble horizon as an infrared cutoff, which has not been explored earlier. Also, we have examined the deviation of the SMHDE model from $\Lambda$CDM using these diagnostic tools. This paper is structured as follows; In Section II, we briefly visit the  Sharma-Mittal holographic dark energy. Section III is dedicated to discussing the flat FRW cosmological model.  Section IV is divided into three subsections A, B and C for the methods of the statefinder hierarchy diagnostic and growth rate of perturbations and diagnostic by the $\omega_{D}-\omega_{D}'$ analysis. Finally, in the last section, we have given inferences.

\section{Sharma-Mittal HDE model}
Recently, inspired by holographic principle and using generalized entropy measure, proposed by Sharma-Mittal \cite{ref30d}, a new form of holographic dark energy model is proposed in \cite{ref38}, called Sharma-Mittal holographic dark energy.\\

By combining the Tsallis and R$\acute{e}$nyi entropies \cite{ref30b,ref30c}, which are two well-known generalized one-parametric entropy measures, with each other, a two-parametric entropy, which was introduced by Sharma-Mittal, is defined in \cite{ref30d}

\begin{equation}
\label{eq1}
S_{SM}=\frac{\left(\frac{A \delta }{4}+1\right)^{R/\delta }-1}{R},
\end{equation}
where $A = 4 \pi L^{2}$ and the IR cutoff is L. Where two free parameters are R and $\delta$. By considering proper limits of R, R$\acute{e}$nyi and Tsallis entropies can be recovered from it. The  Sharma-Mittal entropy becomes R$\acute{e}$nyi entropy in the limit $R \rightarrow 0$, and in limit $R \rightarrow 1 - \delta$, it becomes Tsallis entropy. The energy density is obtained when the UV cutoff and IR cutoff are taken into consideration as was suggested by Cohen et al.\cite{ref57a}.

\begin{equation}
\label{eq2}
\rho _D\propto \frac{S_{SM}}{L^4}\Longrightarrow \rho _D=\frac{3 c^2 S_{SM}}{8 \pi  L^4},
\end{equation}
Considering Hubble horizon cut-off $L = \frac{1}{H}$,  when we take the aforementioned equation into consideration then the energy density of Sharma-Mittal HDE model  \cite{ref38} is 
\begin{equation}
\label{eq3}
\rho _D=\frac{\left(3 c^2 H^4\right) \left(\left(\frac{\pi  \delta }{H^2}+1\right)^{R/\delta }-1\right)}{8 \pi  R},
\end{equation}
where $ c^{2} $ is a numerical constant as usual.

\section{The cosmological model}
For  the flat  FRW  Universe, the metric is given  as :
\begin{eqnarray}
\label{eq4}
ds^{2} = -dt^{2}+a^{2}(t)\Big(dr^{2} + r^{2}d\Omega^{2}\Big)
\end{eqnarray}
In a flat FRW  Universe, the first Friedmann equation, involving dark matter and  SMHDE is defined as :

\begin{eqnarray}
\label{eq5}
H^2 =\frac{1}{3} (8 \pi G) \left(\rho_D+\rho _m\right),
\end{eqnarray}

where $\rho_{D}$ and $\rho_{m}$  represent the energy density of  SMHDE and matter, respectively.   The energy density parameter of  SMHDE and  pressureless matter using the fractional energy densities, can be given as \\
\begin{eqnarray}
\label{eq6}
\Omega_{m} = \frac{8\pi\rho_{m}G}{3H^{2}} , \hspace{1cm} \Omega_{D} = \frac{8\pi\rho_{D}G}{3H^{2}},
\end{eqnarray}

Now Eq. (\ref{eq5}) with help of  Eq. (\ref{eq6}) can be written as:
\begin{eqnarray}
\label{eq7}
1 = \Omega _D+\Omega _m
\end{eqnarray}
The  conservation law for matter and SMHDE  are given as :
\begin{eqnarray}
\label{eq8}
\dot \rho_{m} + 3 H \rho_{m} = 0,
\end{eqnarray}
\begin{eqnarray}
\label{eq9}
\dot \rho_{D} + 3H (\rho_{D} + p_{D}) = 0.
\end{eqnarray}
in which $ \omega _D = p _D/\rho _D$ represents the SMHDE EoS parameter. Now, using differential with time of Eq. (\ref{eq5}) in Eq. (\ref{eq8}), and Eqs. (\ref{eq9}) combined the result with the Eq. (\ref{eq7}), we get\\

$\frac{\dot{H}}{H^2}=-  \left( 3 \left(1-\Omega _D\right) \left(\pi  \delta +H^2\right)\right) \times$
\begin{eqnarray}
\label{eq10}
\frac{1}{2 \left(\pi  c^2 H^2 \left(\frac{\pi  \delta }{H^2}+1\right)^{R/\delta }+\pi  \delta -2 \pi  \delta  \Omega _D-2 H^2 \Omega _D+H^2\right)}
\end{eqnarray}
By Eq. (\ref{eq10}), The deceleration parameter $q$ is found as\\

$q=-1 -\frac{\dot{H}}{H^2}$
\begin{eqnarray}
\label{eq11}
=-1 +\frac{3 \left(\Omega _D-1\right) \left(\pi  \delta +H^2\right)}{2 \left(2 \Omega _D-1\right) \left(\pi  \delta +H^2\right)-2 \pi  c^2 H^2 \left(\frac{\pi  \delta }{H^2}+1\right)^{R/\delta }}
\end{eqnarray}
Now, taking the differential with respect to time of Eq. (\ref{eq3}),  we get
\begin{eqnarray}
\label{eq12}
\dot{\rho _D}=\frac{4 \rho _D \dot H}{H}-\frac{3}{4} c^2 H \dot H \left(\frac{\pi  \delta }{H^2}+1\right)^{\frac{R}{\delta }-1}
\end{eqnarray}

Now by using the Eqs. (\ref{eq12}) with  Eqs. (\ref{eq9}) and (\ref{eq10}), we gets expression for EoS parameter as:\\

$ \omega _D= \Omega _D^{-1}(1 -   \left(\Omega _D-1\right) \left(\pi  \delta +H^2\right) \times $
\begin{eqnarray}
\label{eq13}
\frac{1}{\pi  c^2 H^2 \left(\frac{\pi  \delta }{H^2}+1\right)^{R/\delta }-\left(2 \Omega _D-1\right) \left(\pi  \delta +H^2\right)})
\end{eqnarray}
Also, taking the time differential  of the energy density parameter $\Omega _D $ with Eqs. (\ref{eq10}) and (\ref{eq12}), we find\\

$\Omega '_{D} =- (3 \left(\Omega _D-1\right) (\pi  \left(-c^2\right) H^2 \left(\frac{\pi \delta }{H^2}+1\right)^{R/\delta } $\\

$+ \pi  \delta  \Omega _D+H^2 \Omega _D) \times (  \pi  \left(-c^2\right) H^2  \left(\frac{\pi  \delta }{H^2}+1\right)^{R/\delta } -  $
\begin{eqnarray}
\label{eq14}
\pi  \delta +2 \pi  \delta  \Omega _D+2 H^2 \Omega _D-H^2)^{-1}
\end{eqnarray}

where the dot is the derivative while taking time into consideration and prime lets us obtain the derivative with respect to ln a.\\

\begin{figure}[htp]
	\begin{center}
		\includegraphics[width=8cm,height=8cm]{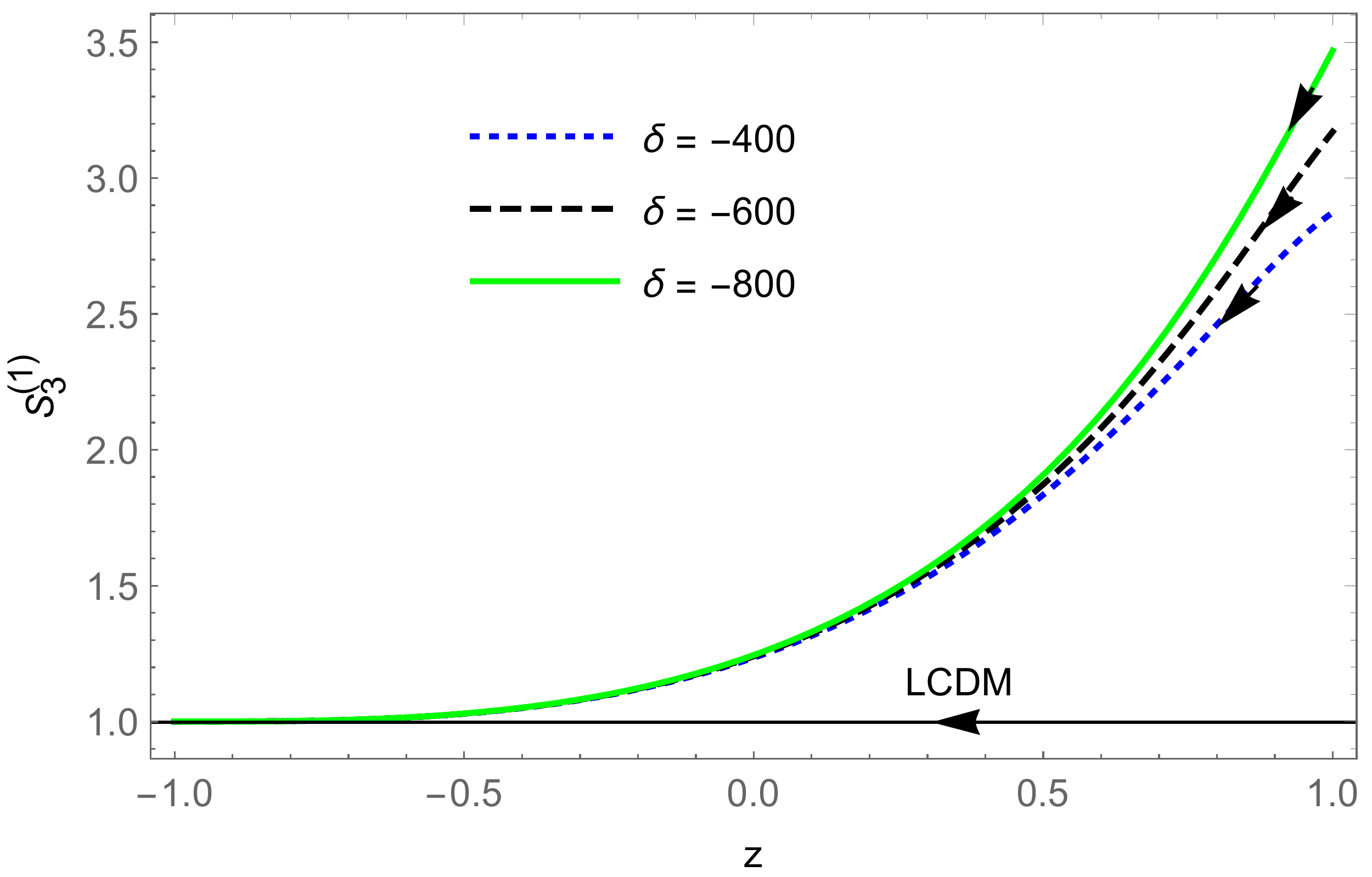}
		\includegraphics[width=8cm,height=8cm]{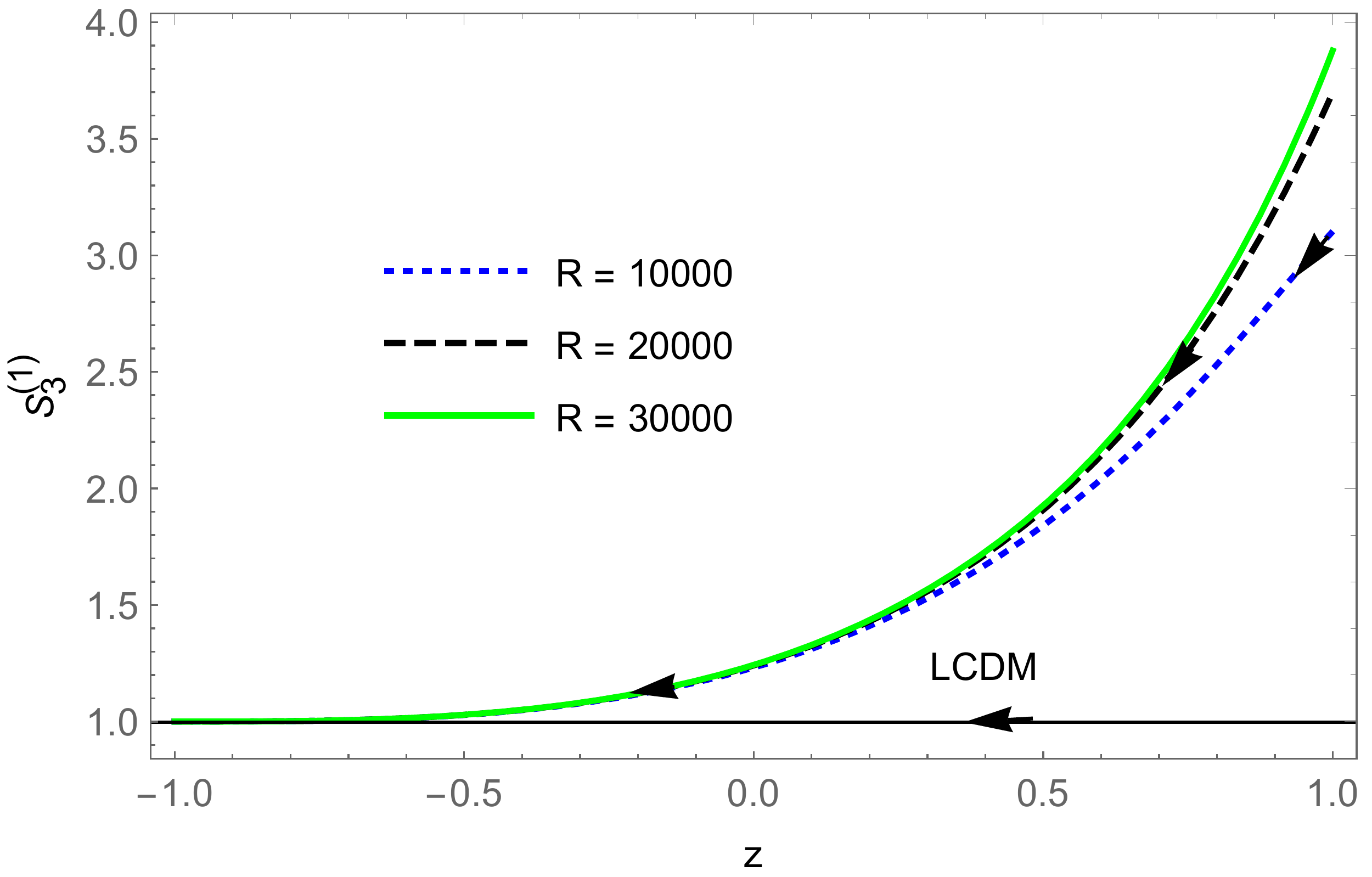}		
		\caption{Graph of $S_3^{(1)}$  versus redshift z, for non- interacting SMHDE with Hbbble radius as the IR cutoff. Here, $H(z=0)= 67$, $\Omega_{m}(z=0) = 0.26$, $R = 10000$ and different 
			values of  $\delta$ (upper panel) and  $H(z=0)= 67$, $\Omega_{m}(z=0) = 0.26$,  $\delta = -600$ and different 
			values of R (below panel).}
		\label{fig:figure1}
	\end{center}
\end{figure}

\section{The methods of diagnostic} 

In this work, we used three diagnostic tools, statefinder hierarchy, the growth rate of perturbations and $\omega_{D}-\omega_{D}^{'}$  pair. We shall explore the SMHDE model to discriminate from the $\Lambda$CDM model with the help of three diagnostic tools in this section.
\subsection{The Statefinder Hierarchy diagnostic}

\begin{figure}	
	\begin{center}
		\includegraphics[width=7cm,height=7cm]{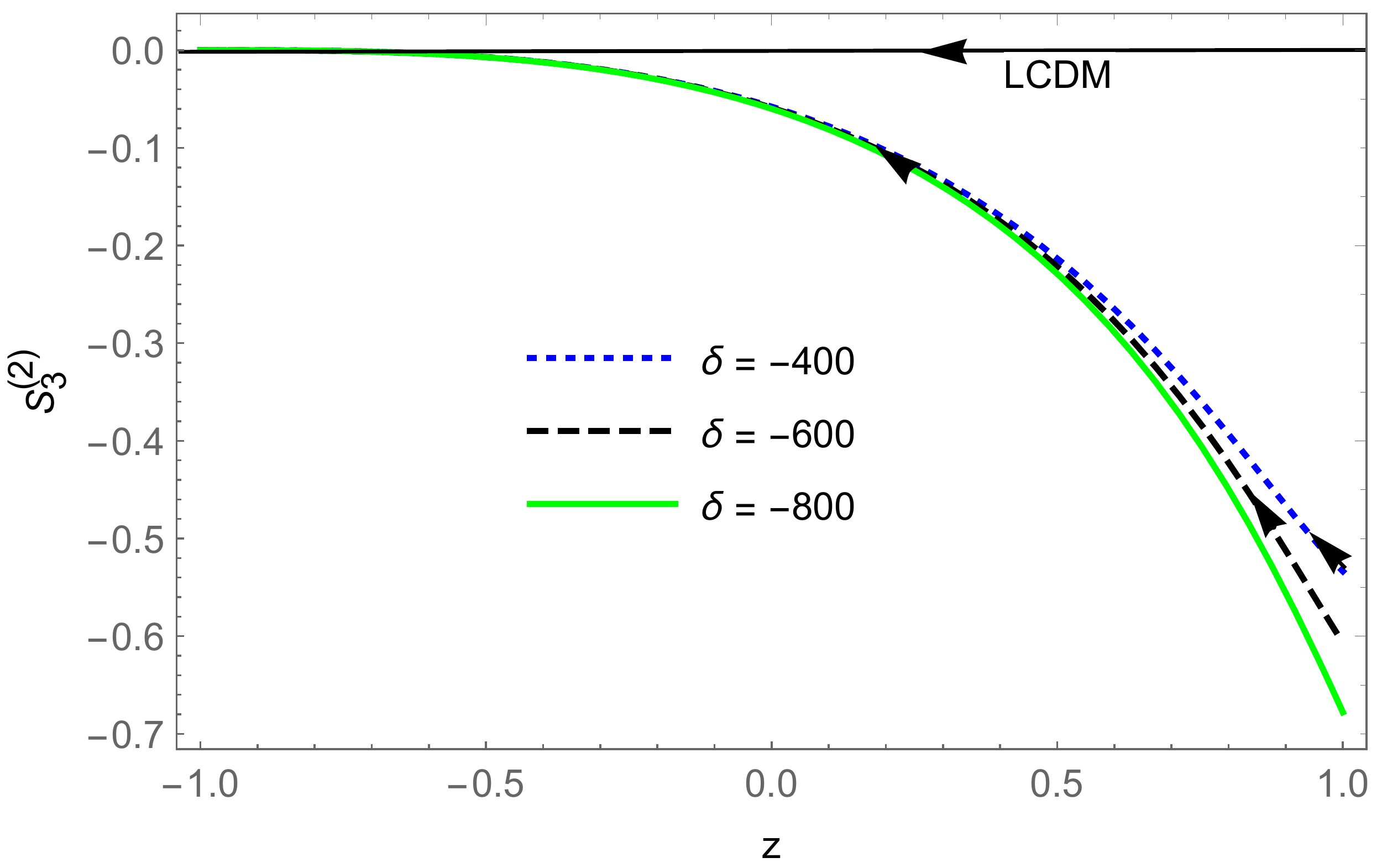}
		\includegraphics[width=7cm,height=7cm]{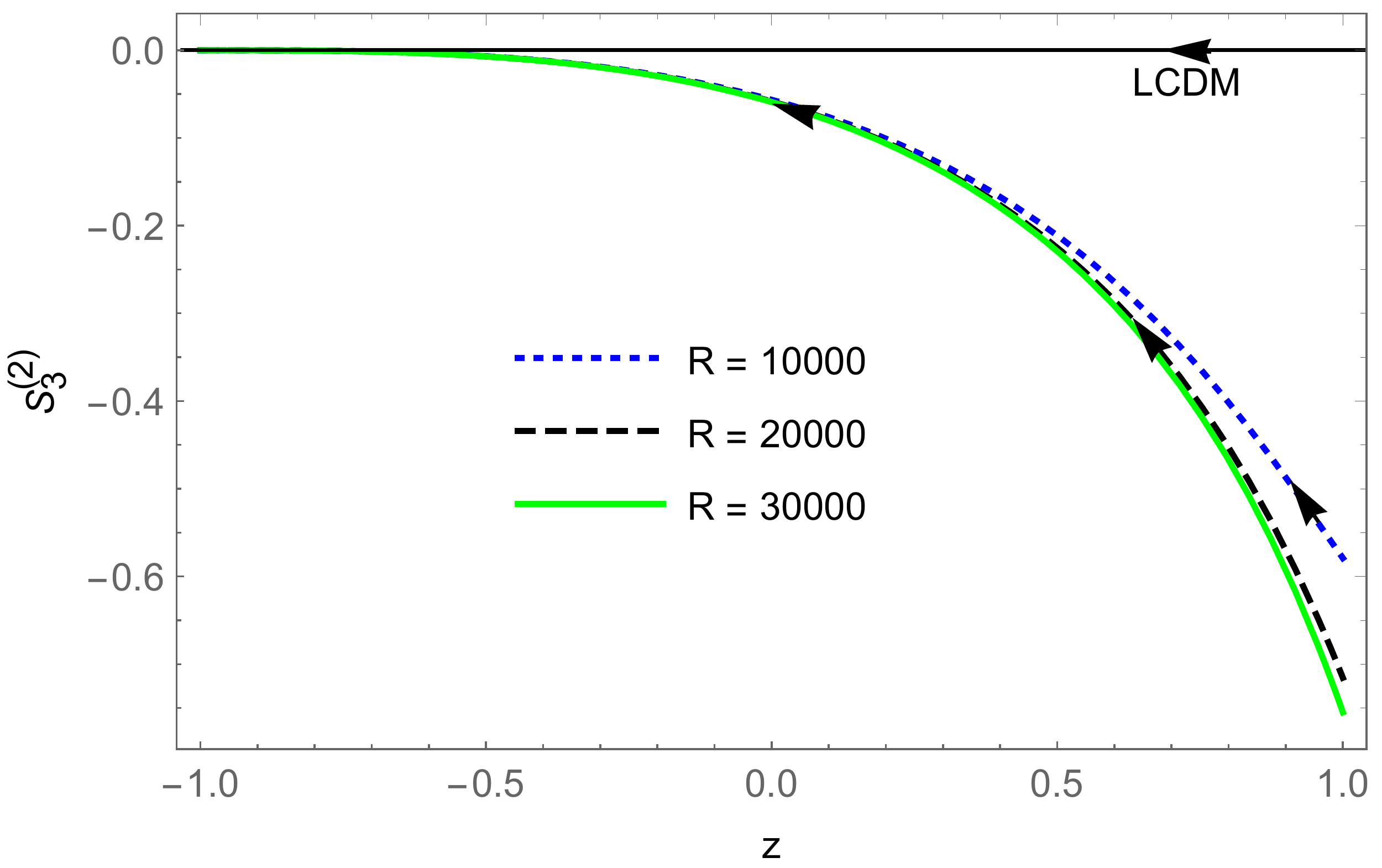}
		\caption {Graph of $S_3^{(2)}$  versus redshift z, for non- interacting SMHDE with Hbbble radius as the IR cutoff. Here, $H(z=0)= 67$, $\Omega_{m}(z=0) = 0.26$, $R = 10000$ and different 
			values of  $\delta$ (upper panel) and  $H(z=0)= 67$, $\Omega_{m}(z=0) = 0.26$,  $\delta = -600$ and different 
			values of R (below panel).}
	\end{center}
	\label{fig:figure2}
\end{figure}

\begin{figure}	
	\begin{center}
		\includegraphics[width=7cm,height=7cm]{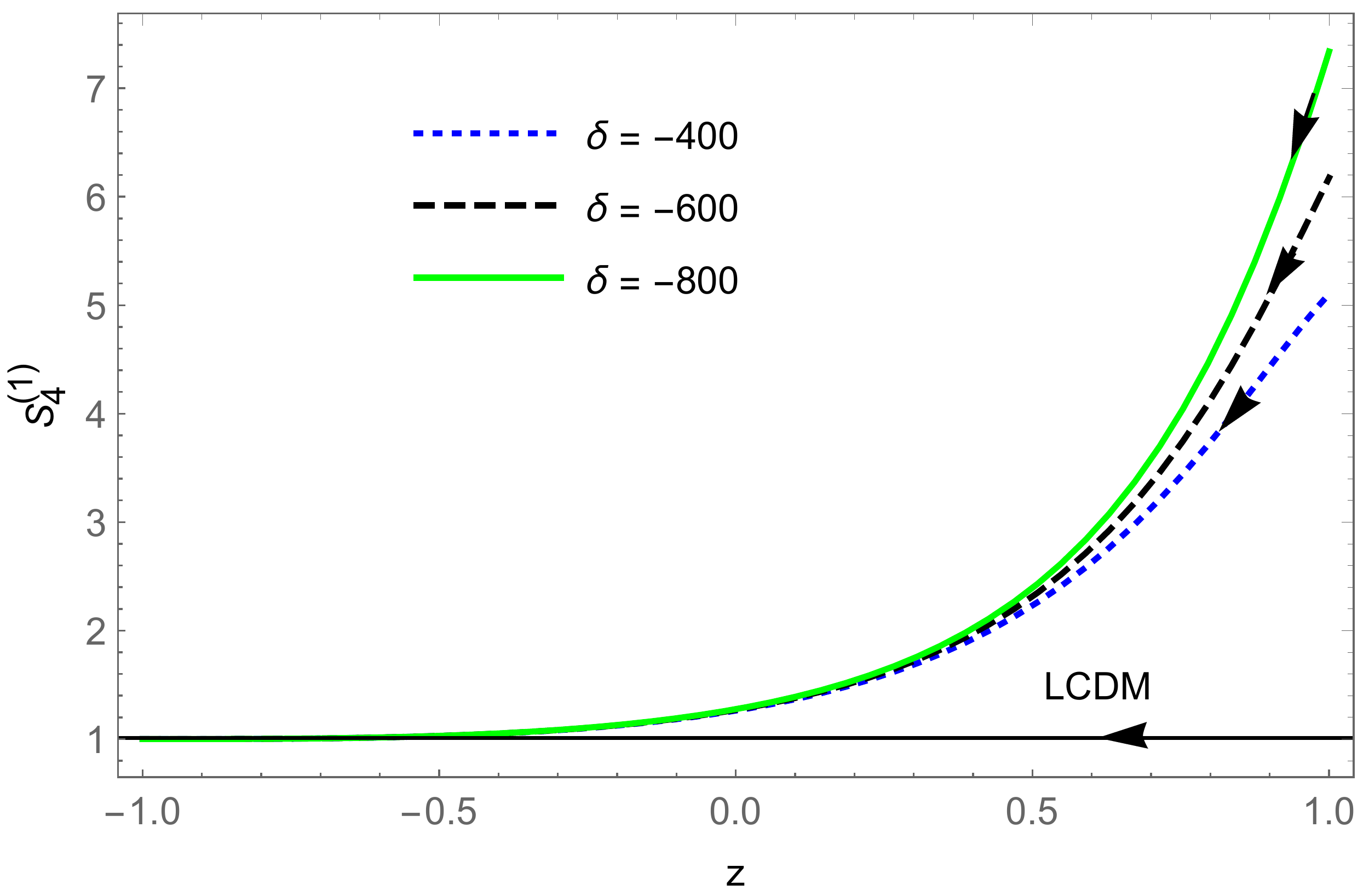}
		\includegraphics[width=7cm,height=7cm]{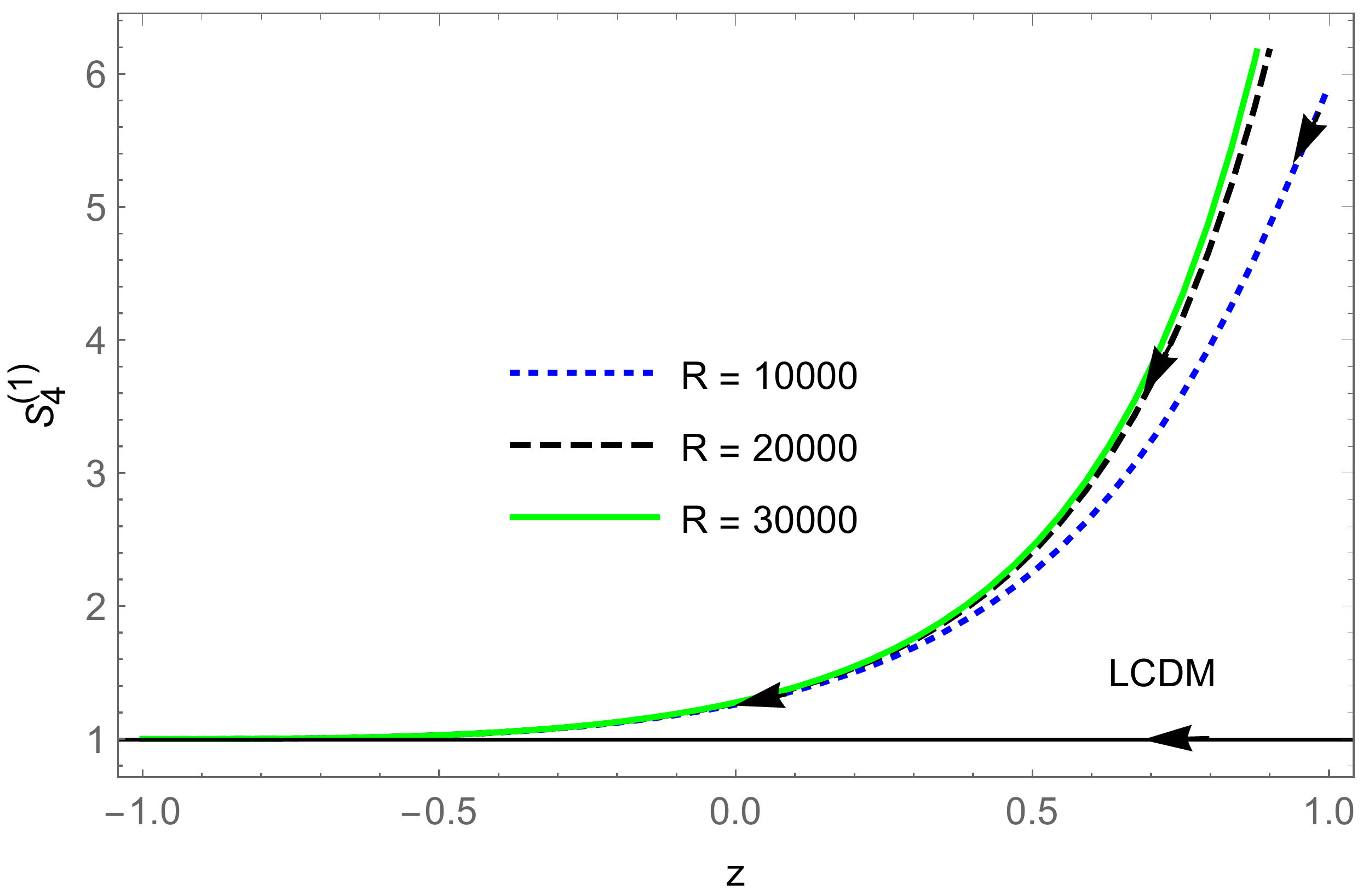}
		\caption {Graph of $S_4^{(1)}$  versus redshift z, for non- interacting SMHDE with Hbbble radius as the IR cutoff. Here, $H(z=0)= 67$, $\Omega_{m}(z=0) = 0.26$, $R = 10000$ and different 
			values of  $\delta$ (upper panel) and  $H(z=0)= 67$, $\Omega_{m}(z=0) = 0.26$  $\delta = -600$ and different 
			values of R (below panel).}
	\end{center}
	\label{fig:figure3}
\end{figure}

\begin{figure}	
	\begin{center}
		\includegraphics[width=7cm,height=7cm]{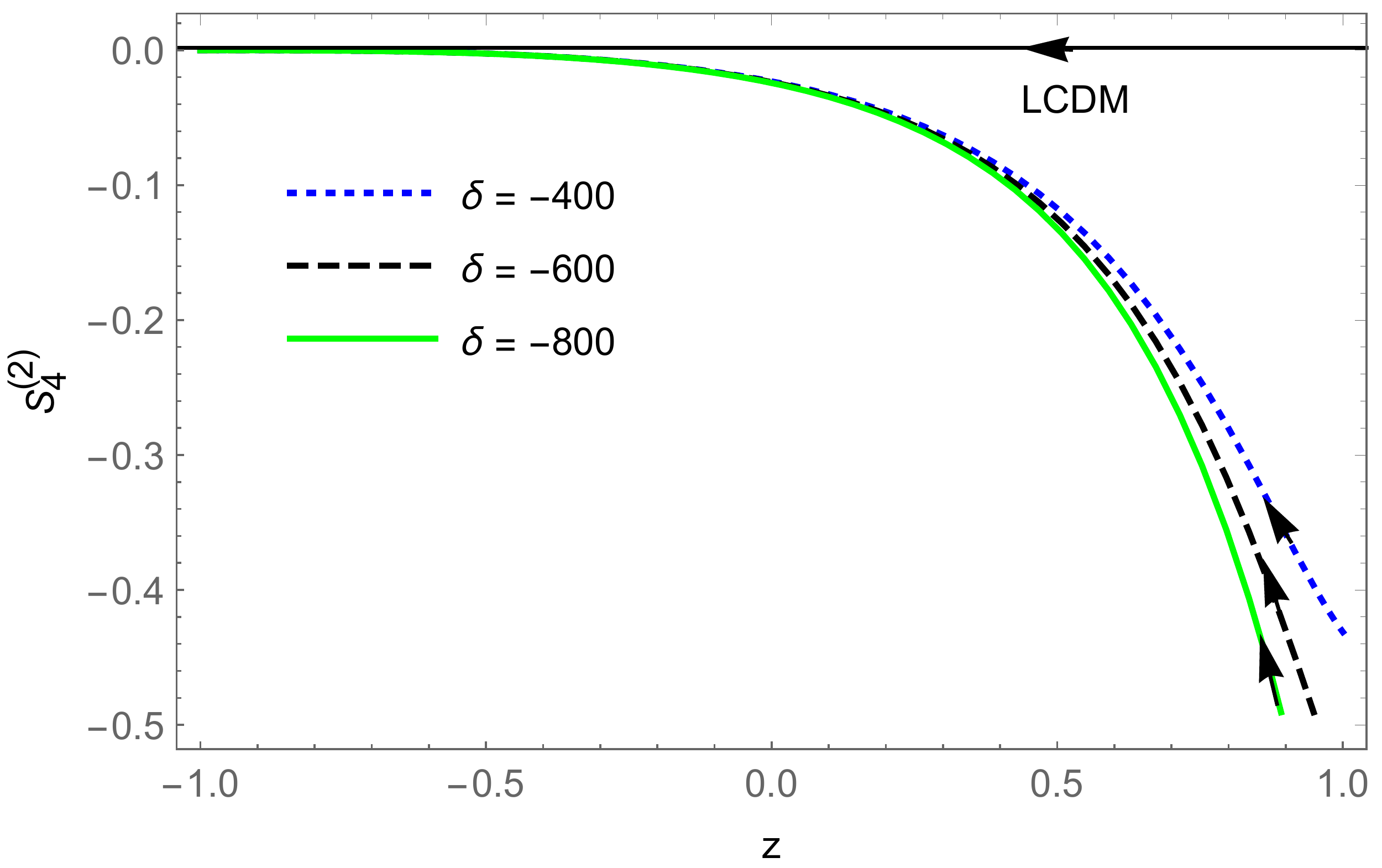}
		\includegraphics[width=7cm,height=7cm]{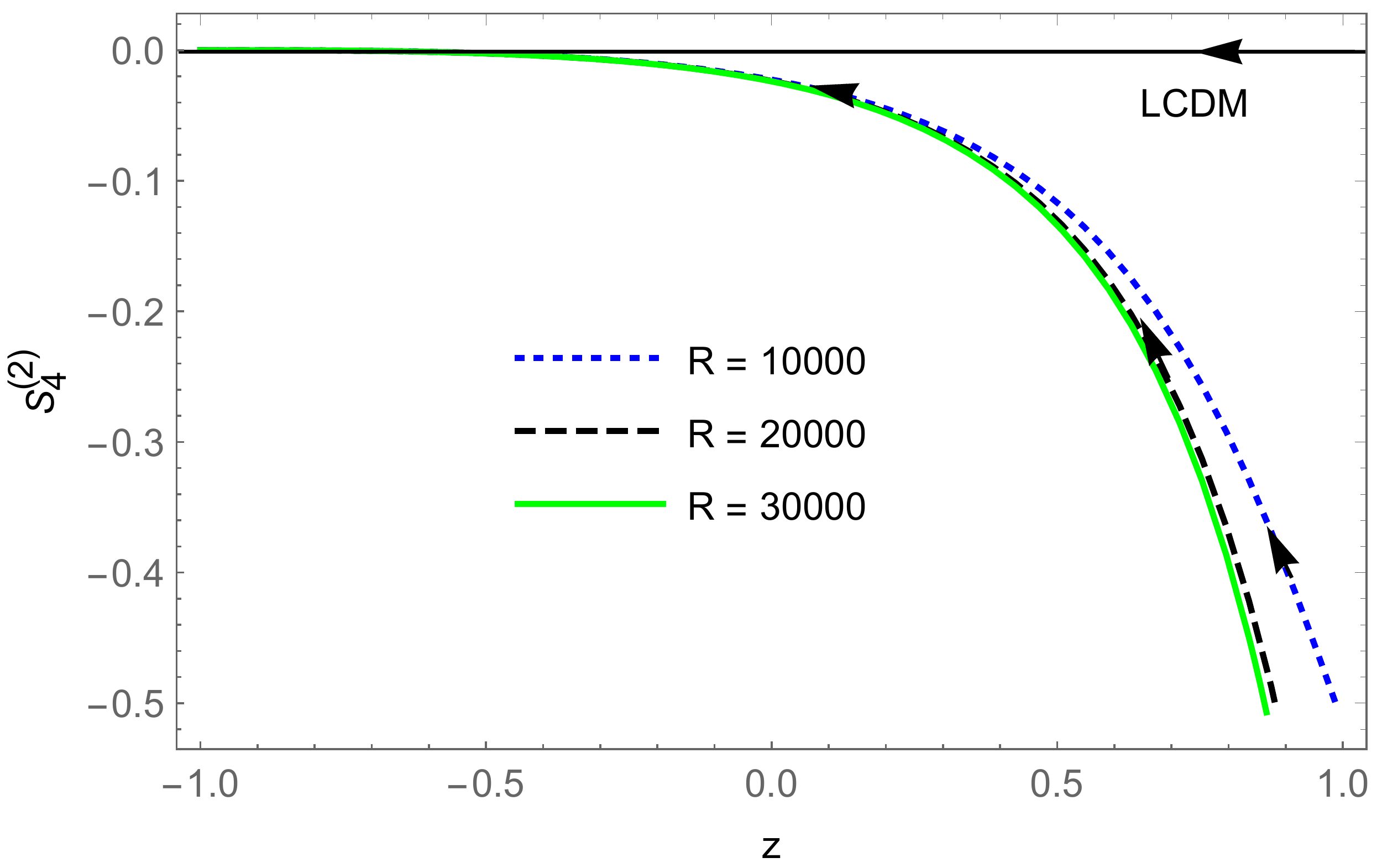}
		\caption {Graph of $S_4^{(2)}$  versus redshift z, for non- interacting SMHDE with Hbbble radius as the IR cutoff. Here, $H(z=0)= 67$, $\Omega_{m}(z=0) = 0.26$, $R = 10000$ and different 
			values of  $\delta$ (upper panel) and  $H(z=0)= 67$, $\Omega_{m}(z=0) = 0.26$,  $\delta = -600$ and different 
			values of R (below panel).}
	\end{center}
	\label{fig:figure4}
\end{figure}

Here, statefinder hierarchy diagnostic will be reviewed and then the growth rate of structure of the SMHDE model will be described. The Taylor expansion of the scale factor $\frac{a(t)}{a_0}=\frac{1}{z+1}$, around the present epoch ${t_0}$ is given as:
\begin{eqnarray}
\label{eq15}
\frac{a (t)}{a_0}=\sum _{n=1}^{\infty } \frac{ A_n(t_0)}{n!}\left[H_0 \left(t-t_0\right)\right]{}^n
\end{eqnarray}

Where $A_n=\frac{a^n}{a H^n}$, ${a^n}$ is the $n^{th}$ derivative of the scale factor a verses cosmic time t and n $\in$ N. The statefinder hierarchy $S_{n}$ is defined as follows \cite{ref87}:
\begin{eqnarray}
\label{eq16}
S_2=A_2+\frac{3 \Omega _m}{2}, S_3=A_3\quad and\quad   S_4=A_4+\frac{9 \Omega _m}{2},
\end{eqnarray}

Aforementioned gives the diagnostics for the model ($\Lambda$CDM) with $n \geq 3$, i.e., $S_{n}|\Lambda$CDM = 1. Hence by the use of $\Omega _m=\frac{2 (q+1)}{3}$ the statefinder hierarchy $S_3^{(1)}$, $S_4^{(1)}$ can be written as:

\begin{eqnarray}
\label{eq17}
S_3^{(1)}=A_3, \quad and \quad S_4^{(1)}=A_4+3 (q+1),
\end{eqnarray}

For $\Lambda$CDM model, $S_{n}^{(1)}$ = 1. In  \cite{ref72} it gives a path for construction of second Statefinder $S_3^{(1)}=S_3$ namely
\begin{eqnarray}
\label{eq18}
S_3^{(2)}=\frac{S_3^{(1)}-1}{3 \left(q-\frac{1}{2}\right)}
\end{eqnarray}

In concordance cosmology $S_3^{(1)} = 1$ while $S_3^{(2) = 0}$. Hence, $\left\{S_3^{(1)},S_3^{(2)}\right\}=\{1,0\}$ gives a model independent
means for forming a distinction between the dark energy models
from the cosmological constant \cite{ref72}. Eq. (\ref{eq18}) gives the second member of the Statefinder hierarchy

\begin{eqnarray}
\label{eq19}
S_n^{(2)}=\frac{S_n^{(1)}-1}{\alpha \left(q-\frac{1}{2}\right)},
\end{eqnarray}
where $\alpha$ is an arbitrary constant. In concordance cosmology $S_n^{(2)}= 0$ and
\begin{eqnarray}
\label{eq20}
\left\{S_n^{(1)},S_n^{(2)}\right\}=\{1,0\},
\end{eqnarray}
Some of degeneracies in $S_n^{(1)}$ can be removed by using the second statefinder $S_n^{(2)}$. For the dark energy model, we have
\begin{eqnarray}
\label{eq21}
S_3^{(1)}=\frac{1}{2} \left(9 \omega _D\right) \left(\omega _D+1\right) \Omega _D+1
\end{eqnarray}
\begin{eqnarray}
\label{eq22}
S_3^{(2)}=\omega _D+1
\end{eqnarray}

$S_4^{(1)} =-\frac{1}{4} \left(27 \omega _D^2\right) \left(\omega _D+1\right) \Omega ^2{}_D -  $\\
\begin{eqnarray}
\label{eq23}
\frac{1}{2} \left(27 \omega _D\right) \left(\omega _D+1\right) \left(\omega _D+\frac{7}{6}\right) \Omega _D+1
\end{eqnarray}
\begin{eqnarray}
\label{eq38}
S_4^{(2)}=-\frac{1}{2} \omega _D \left(\omega _D+1\right) \Omega _D-\left(\omega _D+1\right) \left(\omega _D+\frac{7}{6}\right)
\end{eqnarray}where $S_4^{(2)}=\frac{S_4^{(1)}-1}{9 \left(q-\frac{1}{2}\right)}$ and $q-\frac{1}{2}=\frac{1}{2} \left(3 \omega _D\right) \Omega _D$. As we demonstrate in figures 1, 2, 3, 4 the Statefinder hierarchy $\left\{S_n^{(1)},S_n^{(2)}\right\}$ 
give us a nice way to differenciating  dynamical dark energy models from $\Lambda$CDM model.\\ 

Fig. 1, shows the evolutionary trajectories of $S_3^{(1)}(z)$ for the SMHDE model by considering different values of $\delta$ (upper panel) and $R$ (below panel). Hence,  we investigate two cases. The first is varying $\delta$ with a fixed $R$ (upper panels), the second is varying $R$ and a fixed $\delta$ (below panel). For the evolution of $S_3^{(1)}(z)$ in the SMHDE with varying $R$, the separation of curvilinear shape is not distinct from the SMHDE with varying $\delta$. In the case of varying $\delta$ or $R$ in SMHDE (upper panel or below panel), the curves which are of  $S_3^{(1)}(z)$ have the trajectories on the line of being similar and the trend, which is being followed by of curves $S_3^{(1)}(z)$ is monotonically decreasing at the high-redshift region and then followed by close degeneration together into $\Lambda$CDM $S_3^{(1)} = 1$, at low-redshift region. This shows that different values of $\delta$ or $R$ have quantitative impacts on the $S_3^{(1)}(z)$. Although, in both panels, the curves discriminate well from $\Lambda$CDM in the high-redshift region but highly degenerate in the low-redshift region. \\

Fig. 2, shows the evolutionary trajectories of $S_3^{(2)}(z)$ for the SMHDE model by considering different values of $\delta$ (upper panel) and $R$ (below panel). For the evolution of $S_3^{(2)}(z)$ in the SMHDE with varying R, the differentiation of curvilinear shape is not distinct from the SMHDE with varying $\delta$. In the case of varying $\delta$ or $R$ in the SMHDE (upper panel or below panel), the curves which are of   $S_3^{(2)}(z)$ have the trajectories on the line of being similar and the trend, which is being followed by of curves  $S_3^{(2)}(z)$ is monotonically increasing at the high-redshift region and then followed by close degeneracy together with $\Lambda$CDM  $S_3^{(2)} = 0$, at low-redshift region. This shows that different values of $\delta$ or R have quantitative impacts on the $S_3^{(2)}(z)$. In Fig. 3, we give the graph for $S_4^{(1)}$ evolution versus $z$ i.e. redshift for the SMHDE model by considering different values of $\delta$ (upper panel) and R (below panel). We can say that the evolutionary trajectories of $S_4^{(1)}(z)$ are like that of $S_3^{(1)}(z)$. Quantitative impacts the SMHDE model are found by adopting different values of $\delta$ and $R$.\\

In Fig. 4, we give the graph for  $S_4^{(2)}$ evolution versus $z$ i.e. redshift for the SMHDE model by considering different values of $\delta$ (upper panel) and R (below panel). We can say that the evolutionary trajectories of $S_4^{(2)}$ are like that of  $S_3^{(2)}(z)$. Quantitative impacts on the SMHDE model are found by adopting different values of $\delta$ and $R$ and this is endorsed by the figures.\\

Therefore, in all the plots i.e. Fig. 1-4, there is a drawback that the curves are highly degenerate in the high-redshift region and superposing that of $\Lambda$CDM in the low-redshift region. It means that the single geometric diagnostic is not sufficient. It will be better to combine with the growth rate of perturbations, as CND for getting more clear discrimination.\\

\subsection{Growth rate of perturbations}

\begin{figure}	
	\begin{center}
		\includegraphics[width=7cm,height=7cm]{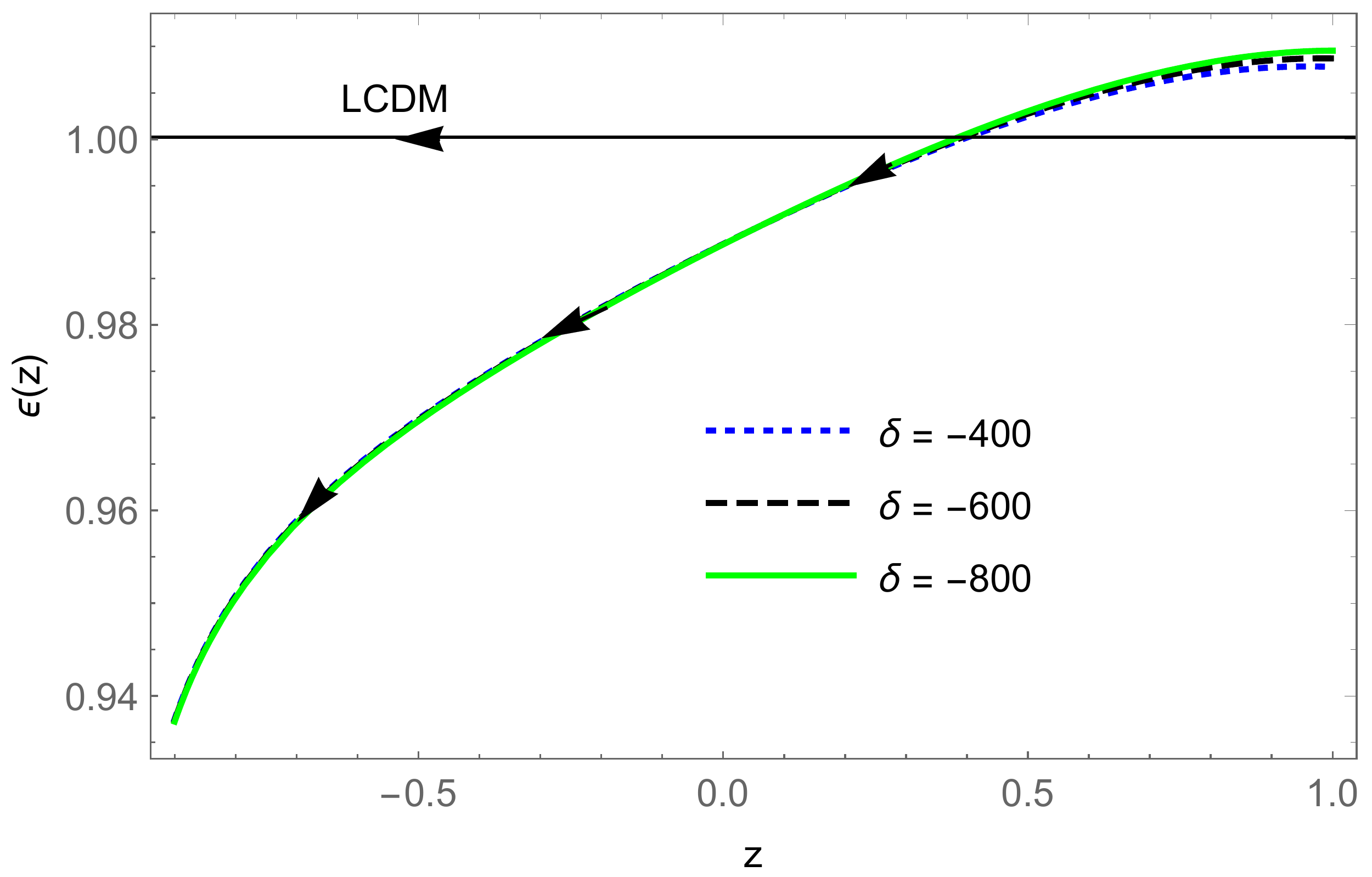}
		\includegraphics[width=7cm,height=7cm]{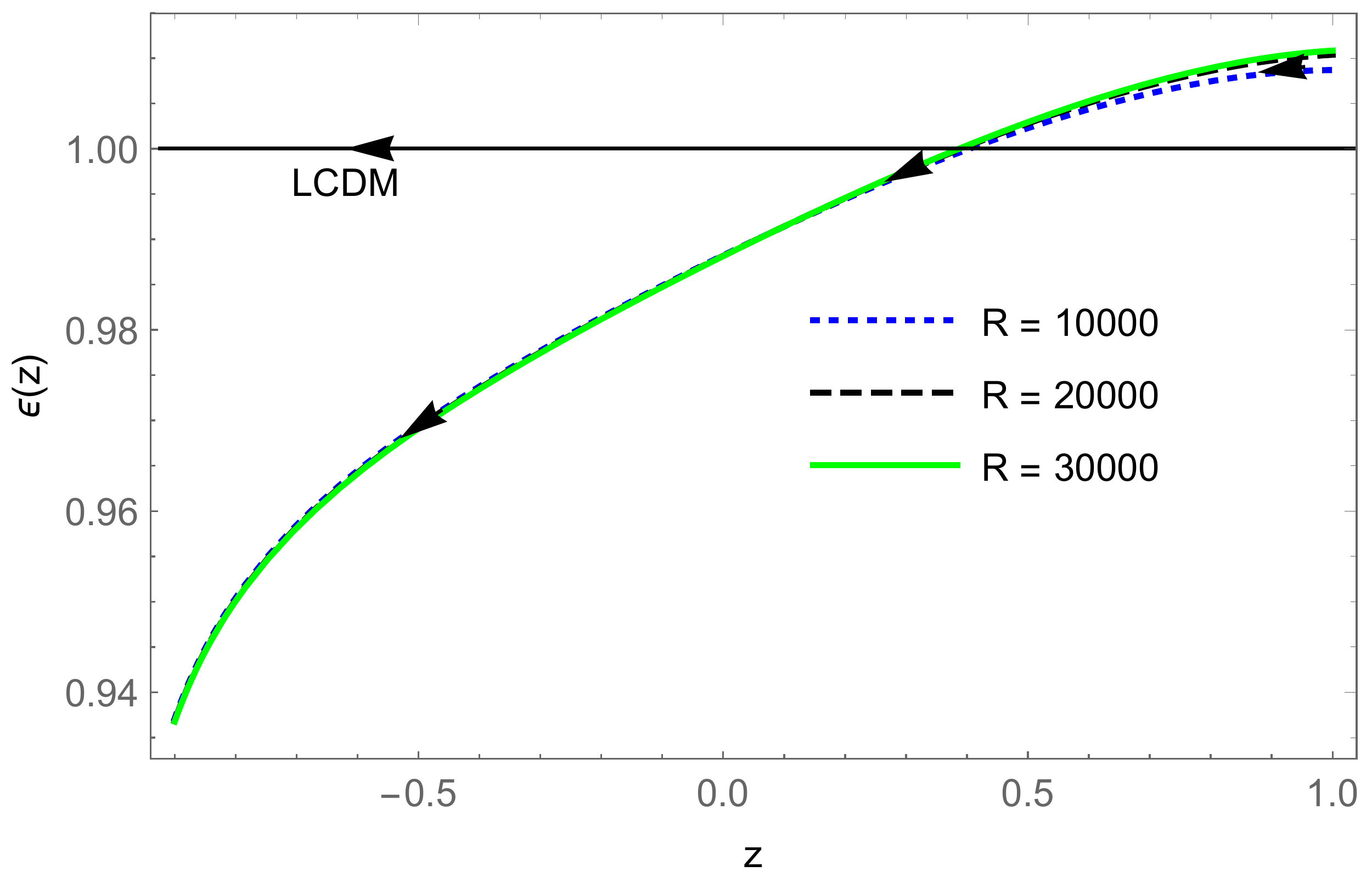}
		\caption {Graph of  $\epsilon(z)$ versus redshift z, for non- interacting SMHDE with Hbbble radius as the IR cutoff. Here, $H(z=0)= 67$, $\Omega_{m}(z=0) = 0.26$, $R = 10000$ and different 
			values of  $\delta$ (upper panel) and  $H(z=0)= 67$, $\Omega_{m}(z=0) = 0.26$,  $\delta = -600$ and different 
			values of R (below panel).}
	\end{center}
	\label{fig:figure5}
\end{figure}
\begin{figure}	
	\begin{center}
		\includegraphics[width=7cm,height=7cm]{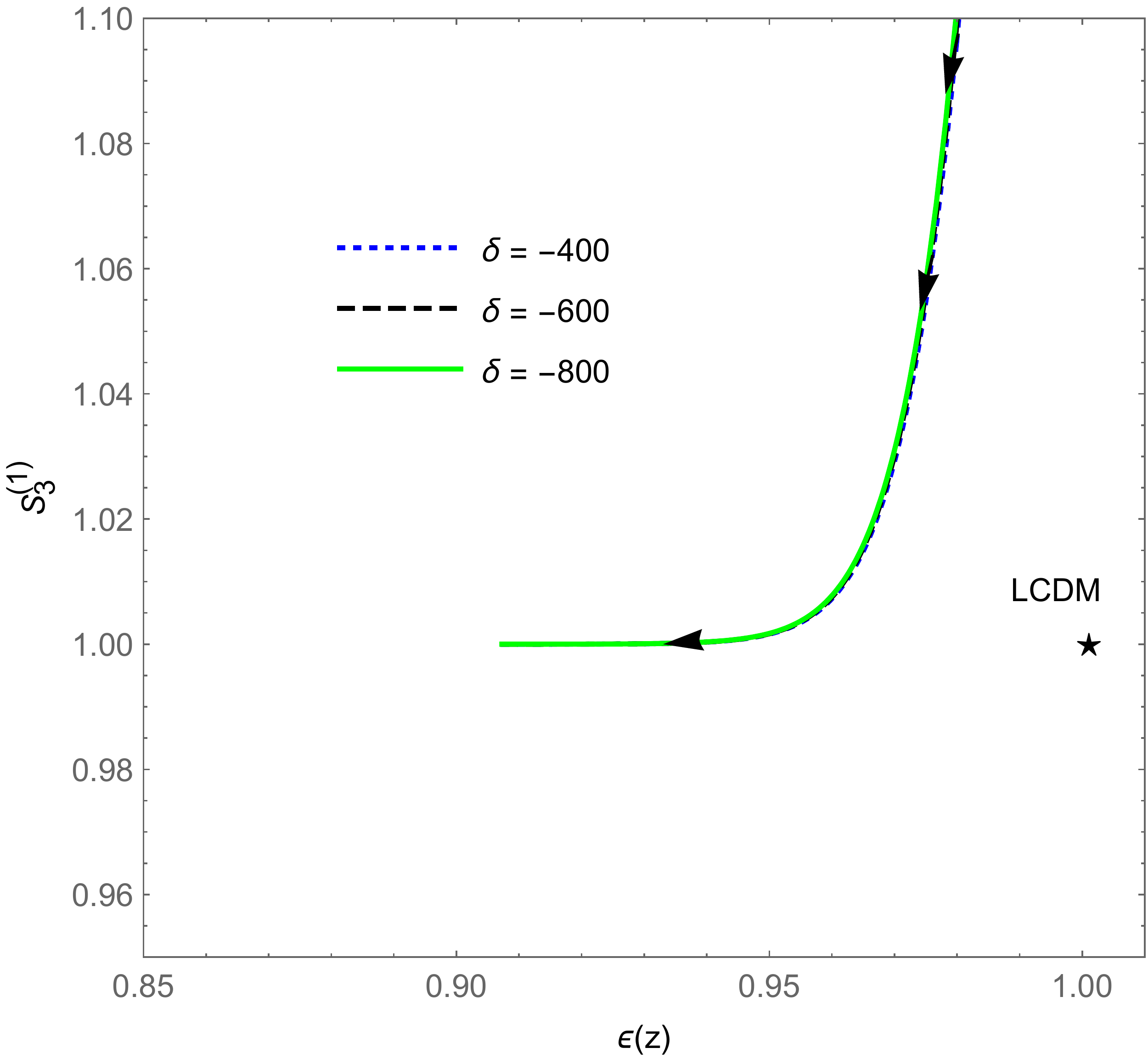}
		\includegraphics[width=7cm,height=7cm]{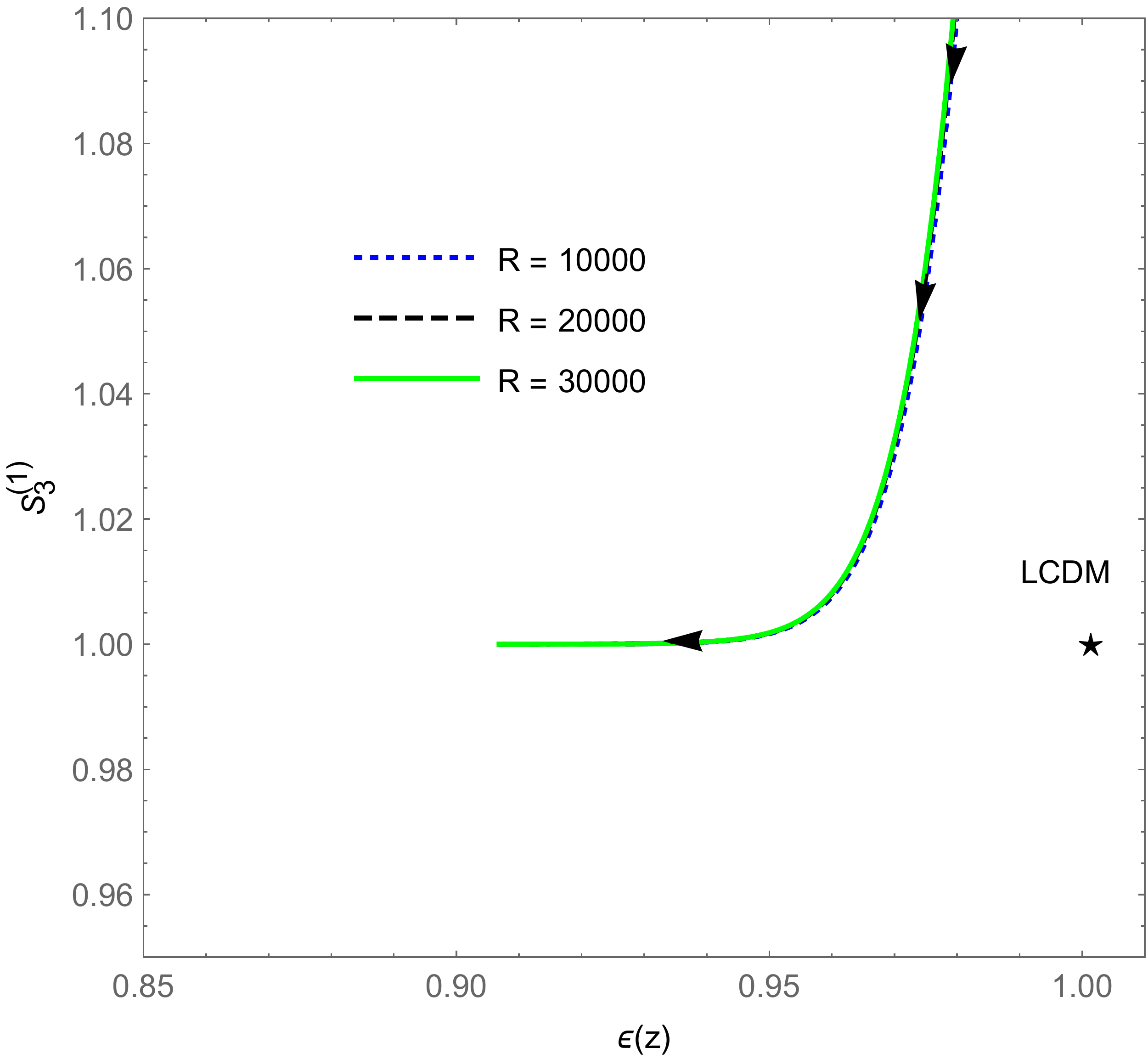}
		\caption {Graph of  Graph of  $S_3^{(1)}$  versus $\epsilon(z)$, for non- interacting SMHDE with Hbbble radius as the IR cutoff. Here, $H(z=0)= 67$, $\Omega_{m}(z=0) = 0.26$, $R = 10000$ and different 
			values of  $\delta$ (upper panel) and  $H(z=0)= 67$, $\Omega_{m}(z=0) = 0.26$,  $\delta = -600$ and different 
			values of R (below panel).}
	\end{center}
	\label{fig:figure6}
\end{figure}

\begin{figure}	
	\begin{center}
		\includegraphics[width=7cm,height=7cm]{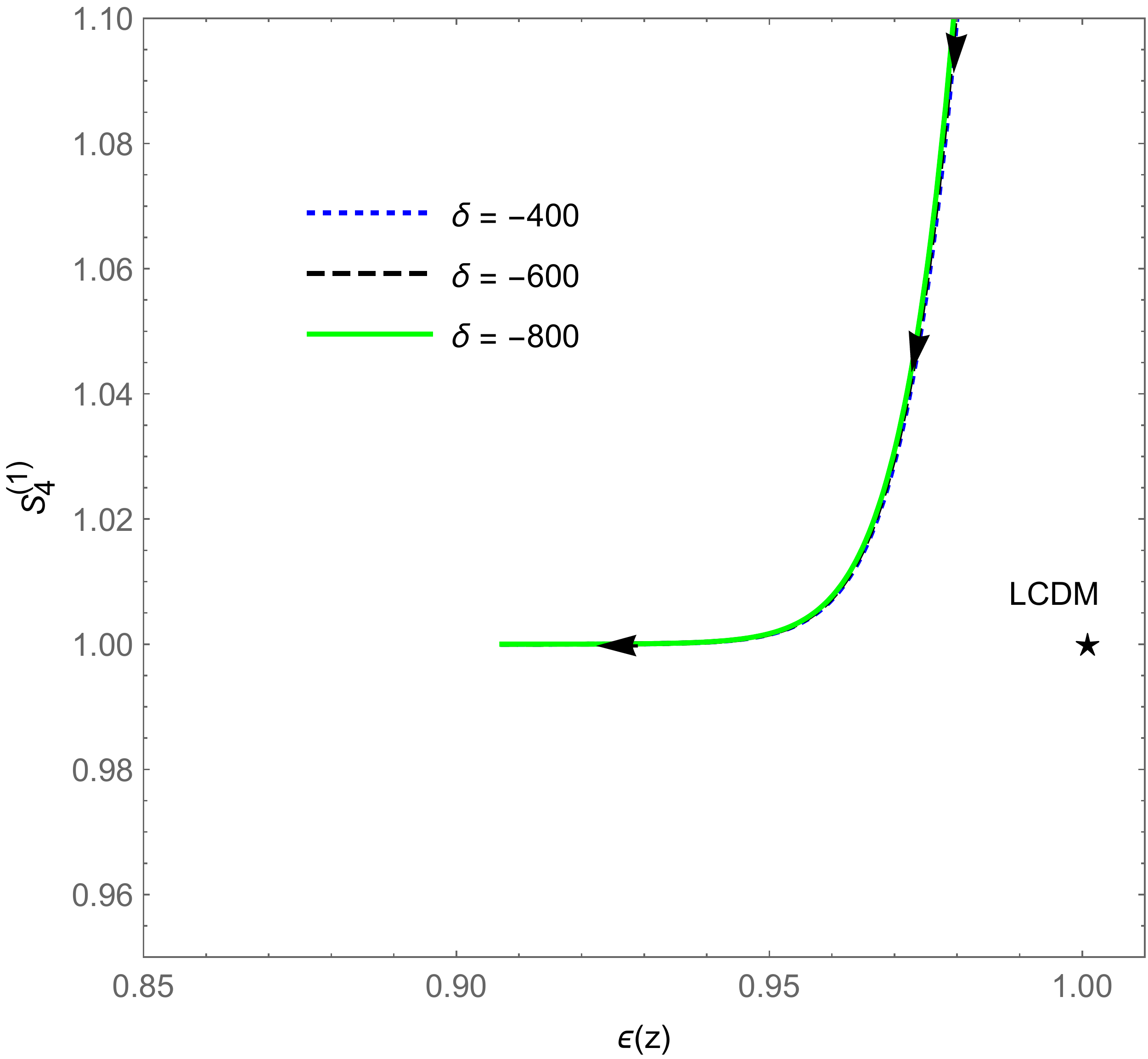}
		\includegraphics[width=7cm,height=7cm]{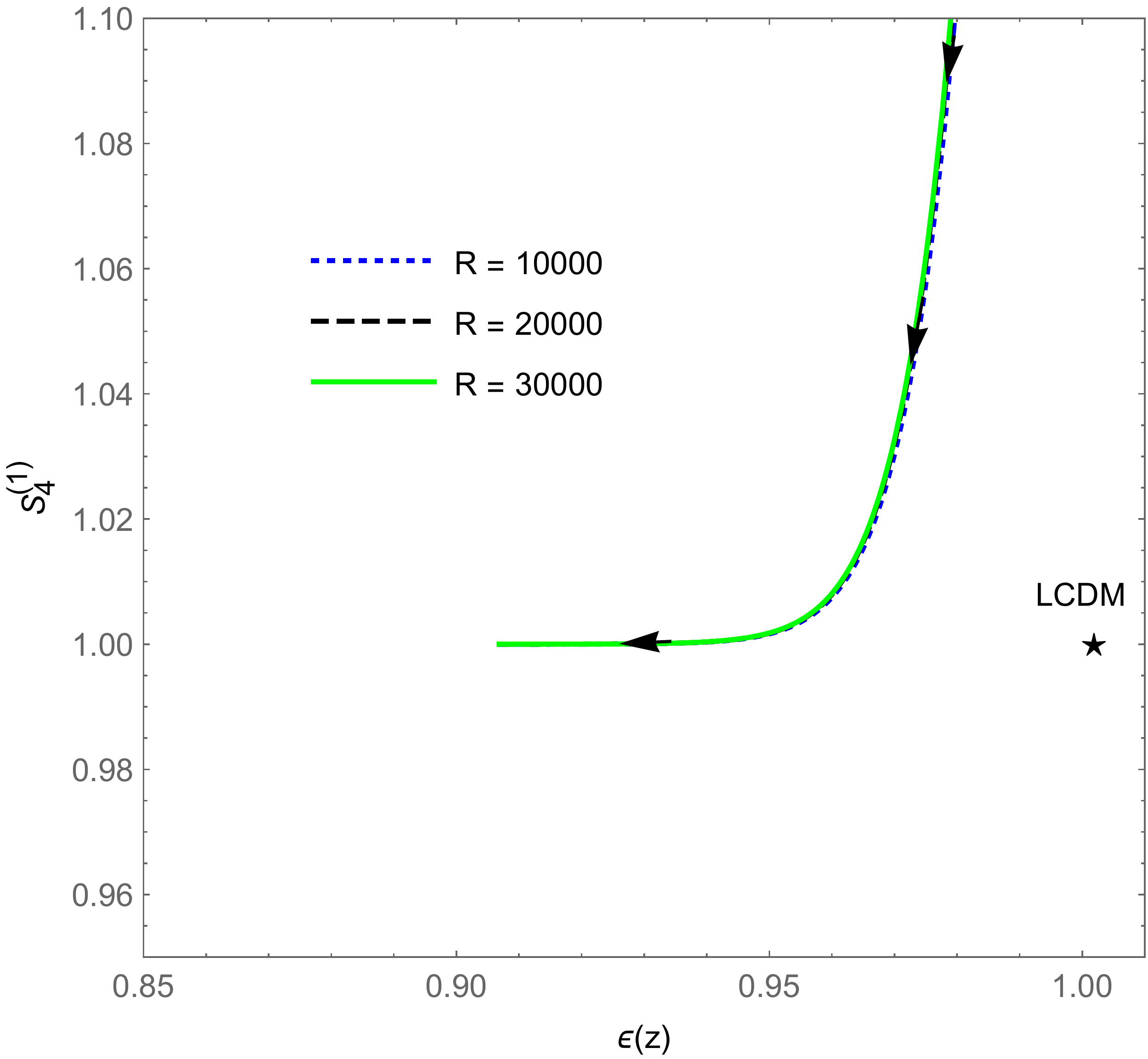}
		\caption {Graph of  Graph of  $S_4^{(1)}$  versus $\epsilon(z)$, for non- interacting SMHDE with Hbbble radius as the IR cutoff. Here, $H(z=0)= 67$, $\Omega_{m}(z=0) = 0.26$, $R = 10000$ and different 
			values of  $\delta$ (upper panel) and  $H(z=0)= 67$, $\Omega_{m}(z=0) = 0.26$,  $\delta = -600$ and different 
			values of R (below panel).}
	\end{center}
	\label{fig:figure6}
\end{figure}

The fractional growth parameter $ \epsilon(z)$  \cite{ref89,ref90} is determined as
\begin{eqnarray}
\label{eq25}
\epsilon(z) =\frac{f (z)}{ f_{\text{$\Lambda $CDM}}(z)}
\end{eqnarray}
Here $f (z)=\frac{d \log \delta }{d \log a }$  is the growth rate of structure. Here, $\delta =\frac{\delta  \rho _m}{\rho _m}$, with $\delta  \rho _m$  and  $\rho _m$ being the the density perturbation and energy density of matter (including CDM and baryons), respectively. If the perturbation is in the linear fashion and without any  interaction between DM and DE, then  we can say that the  equation of perturbation at late times can be:

\begin{eqnarray}
\label{eq26}
\ddot{\delta }+ 2 \dot{\delta } H= 4 \pi  \delta  G \rho _m
\end{eqnarray}

Here, Newton's gravitational constant is represented by $G$. So, the approx  growth rate of linear density perturbation can be reflected by \cite{ref94}:
\begin{eqnarray}
\label{eq27}
f (z)\simeq \Omega _m (z)^{\gamma }
\end{eqnarray}

$\gamma  (z)=\frac{\left(3 \left(1-\omega _D\right) \left(1-\frac{3 \omega _D}{2}\right)\right) \left(1- \Omega _m(z)\right)}{125 \left(1-\frac{6 \omega _D}{5}\right){}^3}$ +
\begin{eqnarray}
\label{eq28}
\frac{3}{5-\frac{\omega _D}{1-\omega _D}}
\end{eqnarray}

where $ \Omega _m(z)=\frac{ \rho _m(z)}{3 H (z)^2 M_p^2}$, the fractional density of matter, $\Omega$ is constant 
or varies slowly with time. $\epsilon(z)$ = 1 and  $\gamma \simeq 0.55$ are the values for the $\Lambda$CDM model \cite{ref94,ref95}. For other models $\epsilon(z)$ exhibits differences from $\Lambda$CDM which would be the possible reason for its use as a diagnostic. By applying the composite null diagnostic ${CND} \equiv \left\{S_{n,}\epsilon \right\}$ where
$ \left\{S_{n,}\epsilon \right\} = \{1,1\}$ for $\Lambda$CDM,  we can make use of both matter perturbational as well as geometrical information of cosmic evolution. While, we can analyze and present only one-side information of cosmic evolution by using one single diagnostic tool. \\

For the diagnose of diverse theoretical DE models, having CND pairs,  $\left\{S_3^{(1)},\epsilon \right\}$ and  $\left\{S_4^{(1)},\epsilon \right\}$, 
the evolution of the fractional growth parameter  $\epsilon (z)$ is analysed.   
Fig. 5 is the evolutionary trajectories of  $\epsilon (z)$ versus redshift $z$ for a spatially homogeneous and an isotropic flat FRW Universe of SMHDE model by considering different values of $\delta$ (upper panel) and $R$ (below panel). For the evolution of $\epsilon (z)$ in the SMHDE with varying $R$, the differentiation of curvilinear shape is not distinct from the SMHDE with varying $\delta$. We can say that the evolutionary trajectories of $\epsilon (z)$ have similar evolutionary trajectories. It is clear from Fig. 5, that the evolutionary trajectories of $\epsilon (z)$ comes closer to $1$ from past to future.\\

The evolutionary trajectories of $\left\{S_3^{(1)},\epsilon \right\}$ of SMHDE model are plotted in Fig. 6 for a spatially homogeneous and an isotropic flat FRW Universe of SMHDE model by considering different values of $\delta$ (upper panel) and $R$ (below panel). The fixed point $(1, 1)$ in this figure presented by  by star symbol denotes the $\Lambda$ CDM. The trend of curves $\left\{S_3^{(1)},\epsilon \right\}$ is monotonically decreasing from  the high-redshift region to low red-shift region for the SMHDE model. This figure clearly detpicts the deviation from $\Lambda$CDM model $\left\{S_3^{(1)}= 1,\epsilon = 1\right\}$ for all values of $\delta$ and $R$ of the SMHDE model. \\

Fig. 7 is the the evolutionary trajectories of the CND pair $\left\{S_4^{(1)},\epsilon \right\}$  for the SMHDE model by considering different values of $\delta$ (upper panel) and R (below panel).
The evolutionary trajectories of  $\left\{S_4^{(1)},\epsilon \right\}$ shows similar characteristic as the curves of $\left\{S_3^{(1)},\epsilon \right\}$. These results shows that adopting different values of $\delta$ and R has quantitative impacts and the deviation from $\Lambda$CDM can be seen in this figure. 

\subsection{The $\omega_{D}-\omega_{D}'$ analysis}
\begin{figure}	
	\begin{center}
		\includegraphics[width=7cm,height=8cm, angle=0]{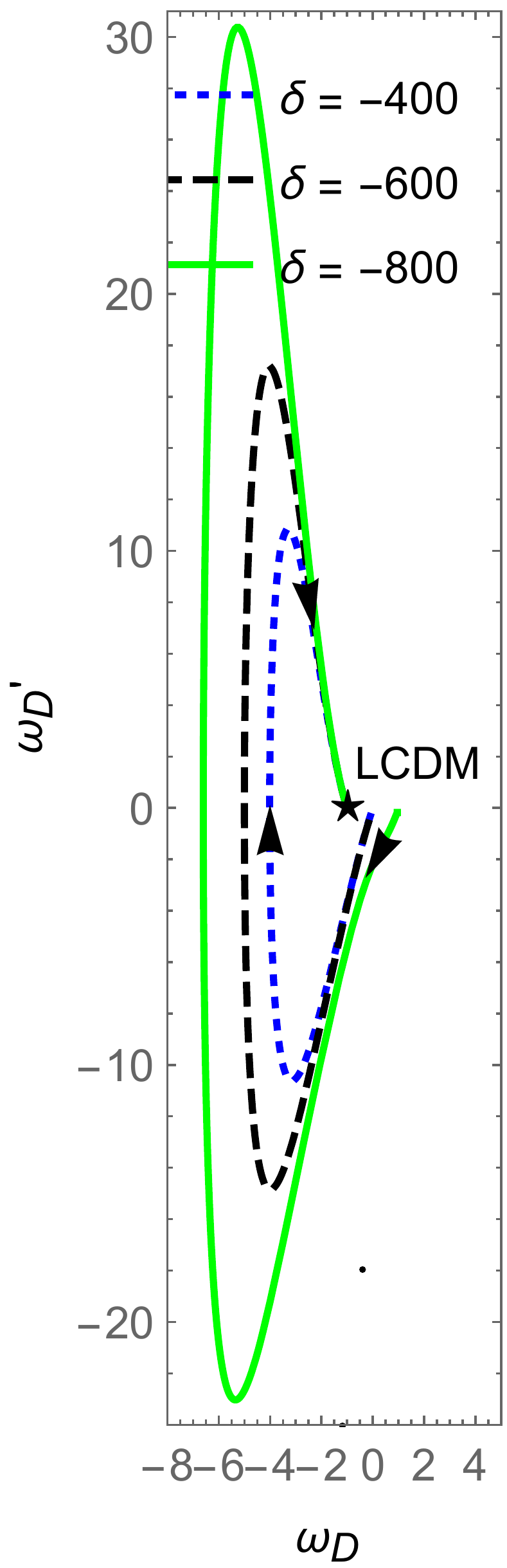}
		\includegraphics[width=7cm,height=7cm, angle=0]{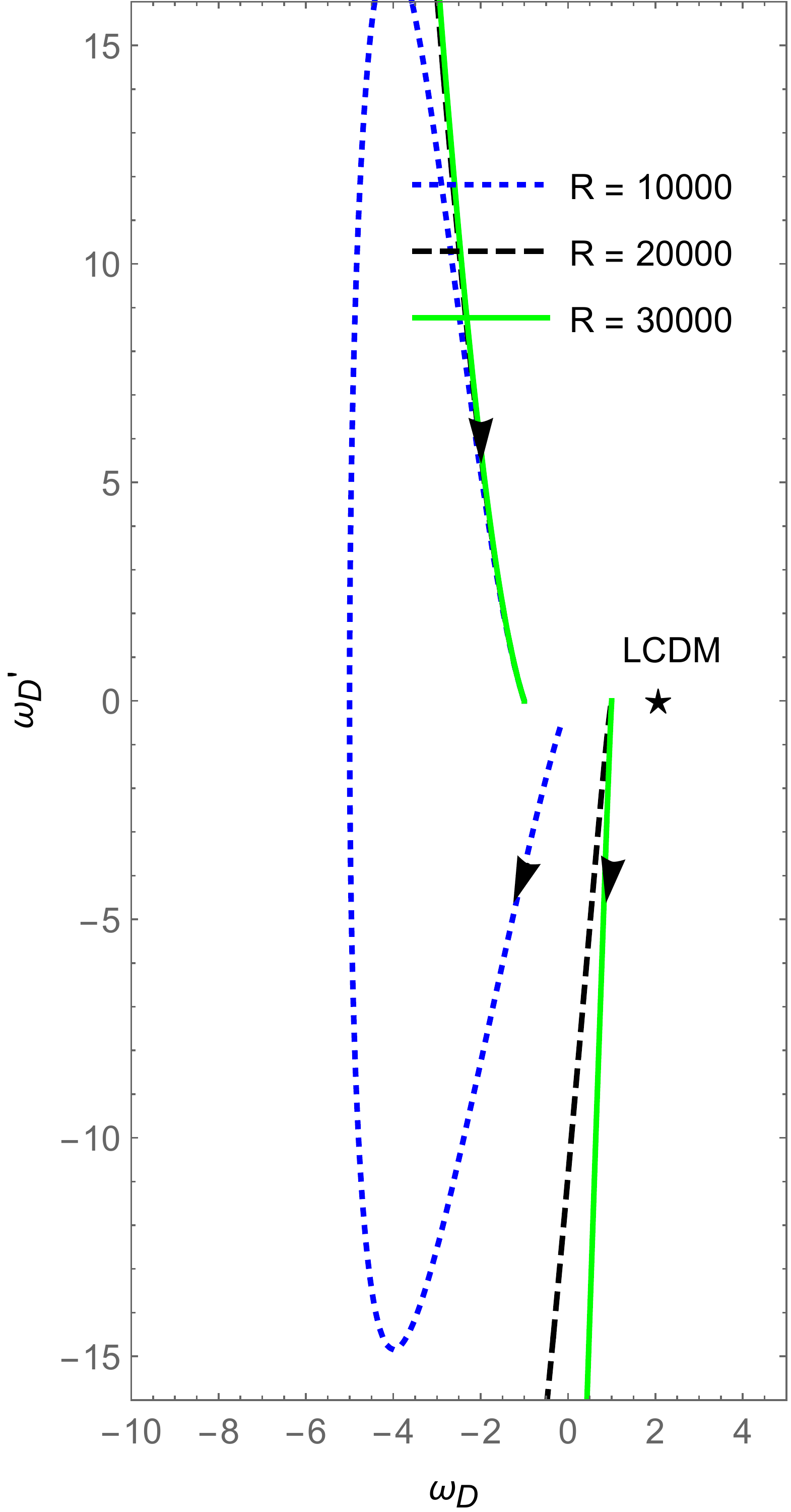} 		
		\caption {The evolution trajectories in the $\omega_{D}- \omega_{D}^{'}$ plane of the SMHDE model, for non- interacting SMHDE with Hbbble radius as the IR cutoff. Here, $H(z=0)= 67$, $\Omega_{m}(z=0) = 0.26$, $R = 10000$ and different 
			values of  $\delta$ (upper panel) and  $H(z=0)= 67$, $\Omega_{m}(z=0) = 0.26$, $\delta = -600$ and different 
			values of R (below panel).}
	\end{center}
	\label{fig:figure7}
\end{figure}

The sign of $\omega_{D}^{'}$ can be used in the thawing and freezing models \cite{ref57} and $\omega_{D}$ is the equation of state  parameter characterizing the dark energy model. Hence  $\omega_{D}- \omega_{D}^{'}$ pair analysis has been used to differentiate the similar model behaviours \cite{ref60,ref61,ref62,ref63,ref64,ref65,ref66}. where  $\omega _D'=\frac{d \omega _D}{ d \log a }$. We investigated the dynamical diagnosis $\omega_{D}- \omega_{D}^{'}$ for SMHDE model which is also utilized widely in the literature. In this dynamical analysis, the fixed point $\omega_{D} = -1$, $\omega_{D}^{'}=0$ represents to the standard  $\Lambda$CDM in the $\omega_{D}-\omega_{D}^{'}$ diagram.\\

The evolutionary trajectories of $\omega_{D}^{'}$ and $\omega_{D}$ plane are shown in Fig. $8$, for an isotropic and spatially homogeneous flat FRW Universe of SMHDE model by considering different values of $\delta$ (upper panel) and $R$ (below panel). It is clear from  $\omega_{D}- \omega_{D}^{'}$ trajectories by considering different values of $\delta$ (upper panel) that for all values of $\delta$  $\omega_{D} \geq -1$ always and it crosses the phantom divide line $\omega_{D} = -1$. It also depicts that currently SMHDE model lies in the thawing region ($\omega_{D}< 0, \omega_{D}^{'}>0$) as well as in freezing region $(\omega_{D}\geq-1)$ which means presently, cosmic expansion is accelerating. 

\section{Conclusions}

The paper uses the Sharma-Mittal Holographic Dark Energy (SMHDE) model in flat FRW  Universe by considering different Sharma-Mittal parameter $\delta$. This can be summarized as \\
\begin{itemize}

	\item 
	We studied the deviation of SMHDE model from  $\Lambda$CDM regarding  different values of Sharma-Mittal parameter $\delta$ by the use of the diagnostics of statefinder hierarchy and growth rate of structure. The statefinder hierarchy gives analytical expressions of $S_3^{(1)}$, $S_3^{(2)}$, $S_4^{(1)}$ and $S_4^{(2)}$, for SMHDE in as cosmological parameters. To check the growth rate of structure $\epsilon (z)$ has been calculated analytically for the SMHDE model. We tested the SMHDE model using $\left\{S_3^{(1)},\epsilon \right\}$ diagnostics. We plotted the evolution curves of $S_3^{(1)}$, $S_3^{(2)}$, $S_4^{(1)}$ and $S_4^{(2)}$ with respect to cosmic time $z$ and $\epsilon (z)$. These evolutionary trajectories shows that the SMHDE model shows $\Lambda$CDM behaviour at late time. We have plotted the evolutionary trajectories of  $\left\{S_3^{(1)},\epsilon \right\}$ plane which depicts that SMHDE model for all values of $\delta$ shows same deviation from $\Lambda$CDM model.\\
	
	\item The various diagnostic methods for dark energy have been discussed. we have also examined the $\omega_{D}-\omega_{D}'$ pair analysis for our SMHDE model in subsection $C$. These analysis are used to differentiate among various dark energy models.  In the subsection $5.1$ we investigated dynamical diagnosis $\omega_{D}- \omega_{D}^{'}$ for SMHDE model where the derivative with respect to log a is denoted by prime notation. The evolutionary trajectories of $\omega_{D}- \omega_{D}^{'}$ shows that presently cosmic expansion is accelerating since our SMHDE model lies in the thawing region ($\omega_{D}< 0, \omega_{D}^{'}>0$). It also indicates that $\omega_{D}$  crosses the phantom divide line $\omega_{D} = -1$ and $\omega_{D} \geq -1$. \\
	
	We have used three diagnostic tools in this work such as $\omega_{D}-\omega_{D}'$ pair, the growth rate of perturbations and statefinder hierarchy to diagnose the SMHDE model. Some other diagnostic tool like statefinder diagnostic can also be used to discriminate the SMHDE from $\Lambda$CDM model.\\
	
	We hope that in future high precision observations, for example, SNAP-type investigation can be equipped for deciding the cosmological parameters exactly and consequently identify the correct cosmological model and closer to understand the properties of the SMHDE model.
\end{itemize}

\section*{Acknowledgments}
The authors are thankful for valuable suggestions given by Dr. Prateek Pandey, GLA University, Mathura, India, in this research work.

\end{document}